\documentclass[sigconf,10pt]{acmart}

\usepackage[english]{babel}
\usepackage{blindtext}
\usepackage{tikz}
\usepackage{pifont}
\usepackage{algpseudocode}
\usepackage{algorithm}
\usepackage{multirow}
\usepackage{svg}
\usepackage[font=small,format=plain,labelfont=bf,textfont=bf]{caption}
\usepackage{enumerate}
\usepackage{paralist}
\usepackage{enumitem}[noitemsep,topsep=0pt,leftmargin=0pt]
\setitemize{noitemsep,topsep=0pt,parsep=0pt,partopsep=0pt}
\setenumerate{noitemsep,topsep=0pt,parsep=0pt,partopsep=0pt}
\usepackage[capitalize]{cleveref}
\crefname{section}{\S}{\S\S}
\usepackage{tcolorbox}

\newif\ifsubmission
\submissiontrue

\ifsubmission
\newcommand{\mcnote}[1]{}
\else
\usepackage[textsize=tiny]{todonotes}
\newcommand{\mcnote}[1]{\todo[color=orange!40,inline]{marco: #1}} 
\fi
\newcommand{\AIsolate}{\textsc{HarmonAIze}\xspace}
\newcommand{\smartparagraph}[1]{\noindent{\bf #1}\ }

\begin{document}

\acmYear{2025}\copyrightyear{2025}
\setcopyright{cc}
\setcctype[4.0]{by}
\acmConference[APSys '25]{16th ACM SIGOPS Asia-Pacific Workshop on Systems}{October 12--13, 2025}{Seoul, Republic of Korea}
\acmBooktitle{16th ACM SIGOPS Asia-Pacific Workshop on Systems (APSys '25), October 12--13, 2025, Seoul, Republic of Korea}
\acmDOI{10.1145/3725783.3764395}
\acmISBN{979-8-4007-1572-3/25/10}



\title{Cloud abstractions for AI workloads}
\author{Marco Canini}
\affiliation{%
  \institution{KAUST}%
  \city{}
  \country{}%
}
\author{Theophilus A. Benson}
\affiliation{%
  \institution{Carnegie Mellon University}%
  \city{}
  \country{}%
}
\author{Ricardo Bianchini}
\affiliation{%
  \institution{Microsoft}%
  \city{}
  \country{}%
}
\author{Íñigo Goiri}
\affiliation{%
  \institution{Microsoft}%
  \city{}
  \country{}%
}
\author{Dejan Kostić}
\affiliation{%
  \institution{KTH Royal Institute of Technology}%
  \city{}
  \country{}%
}
\author{Peter Pietzuch}
\affiliation{%
  \institution{Imperial College London}%
  \city{}
  \country{}%
}
\author{Simon Peter}
\affiliation{%
  \institution{University of Washington}%
  \city{}
  \country{}%
}

\date{}

\begin{abstract}

AI workloads, often hosted in multi-tenant cloud environments, require vast computational resources but suffer inefficiencies due to limited tenant-provider coordination. Tenants lack infrastructure insights, while providers lack workload details to optimize tasks like partitioning, scheduling, and fault tolerance. We propose \AIsolate to redefine cloud abstractions, enabling cooperative optimization for improved performance, efficiency, resiliency, and sustainability. We outline key opportunities and challenges this vision faces.

\end{abstract}

\maketitle
\renewcommand{\shortauthors}{Canini et al.}

\mcnote{Mesos management layer for fine-grained resource sharing}

\section{Introduction}

\mcnote{iterative nature of AI workloads as compared to general distributed systems}
\mcnote{AI are acccelerated workloads. not good at control flow. easier to predict performance.}

\mcnote{what the baseline looks like. this is similar to peter's comment on the need for a table to give a few numbers.}

\mcnote{figure 1. ricardo's 20 comment on the figure.}

Modern AI workloads are widely compute-intensive, driven by the growing adoption of generative AI (GenAI), agentic applications, and large language models (LLMs). These workloads require large clusters of specialized hardware accelerators, such as GPUs and TPUs, to efficiently train, fine-tune or serve models that contain billions or even trillions of parameters. 
Training and serving state-of-the-art AI models consumes substantial amounts of power, contributing to carbon emissions and raising sustainability challenges~\cite{wu2024sustainableai,wu2022sustainableai}. Optimizing the efficiency of AI workloads is essential to mitigate their environmental impact.
As these workloads scale to meet growing demands, their computational requirements outpace traditional hardware capabilities, necessitating not only energy-efficient algorithms~\cite{chung2024perseus,you2023zeus} but also innovative infrastructure designs that maximize resource utilization.

To maximize returns on costly hardware accelerators (e.g., GPUs costing tens of thousands of dollars), AI workloads are increasingly hosted in multi-tenant cloud environments. These environments enable resource sharing, offering tenants scalability, flexibility, and cost-efficiency while minimizing idle hardware and maximizing infrastructure utilization.

However, tenants operating in these environments face a fundamental challenge: they lack visibility into the infrastructure details (such as network topology, job placement, resource availability, etc.) that could be used to optimize their workloads. Providers often withhold such information to maintain commercial competitiveness or to preserve the flexibility to make changes to their infrastructure. This lack of awareness creates a barrier to fine-grained optimizations that could benefit both tenants and providers.

Providers are ideally placed to optimize infrastructure-level tasks critical to AI workloads, such as collectives, job scheduling and scaling, and fault tolerance~\cite{wu2024mccs,he2024unicron,jiang2024megascale,duan2024parcae}. Yet, without insights into tenant workloads -- such as communication patterns~\cite{zheng2022alpa,li2022amp} or hyperparameter adaptations~\cite{mai2020kungfu,qiao2021pollux} -- providers cannot optimize these tasks effectively.

The root cause of these inefficiencies lies in the lack of abstractions enabling cooperative optimization between tenants and providers (\cref{sec:background}), a challenge shared across AI and other cloud applications~\cite{huang2024workloadintelligence,bilal2023serverless,chen2022nethint}. Such abstractions would need to provide semantic information necessary for cross-layer control, allowing tenants and providers to jointly optimize performance, efficiency, availability and resilience, achieving outcomes that neither could accomplish alone.

\begin{figure}[t!]
    \centering
    \includegraphics[width=0.76\columnwidth]{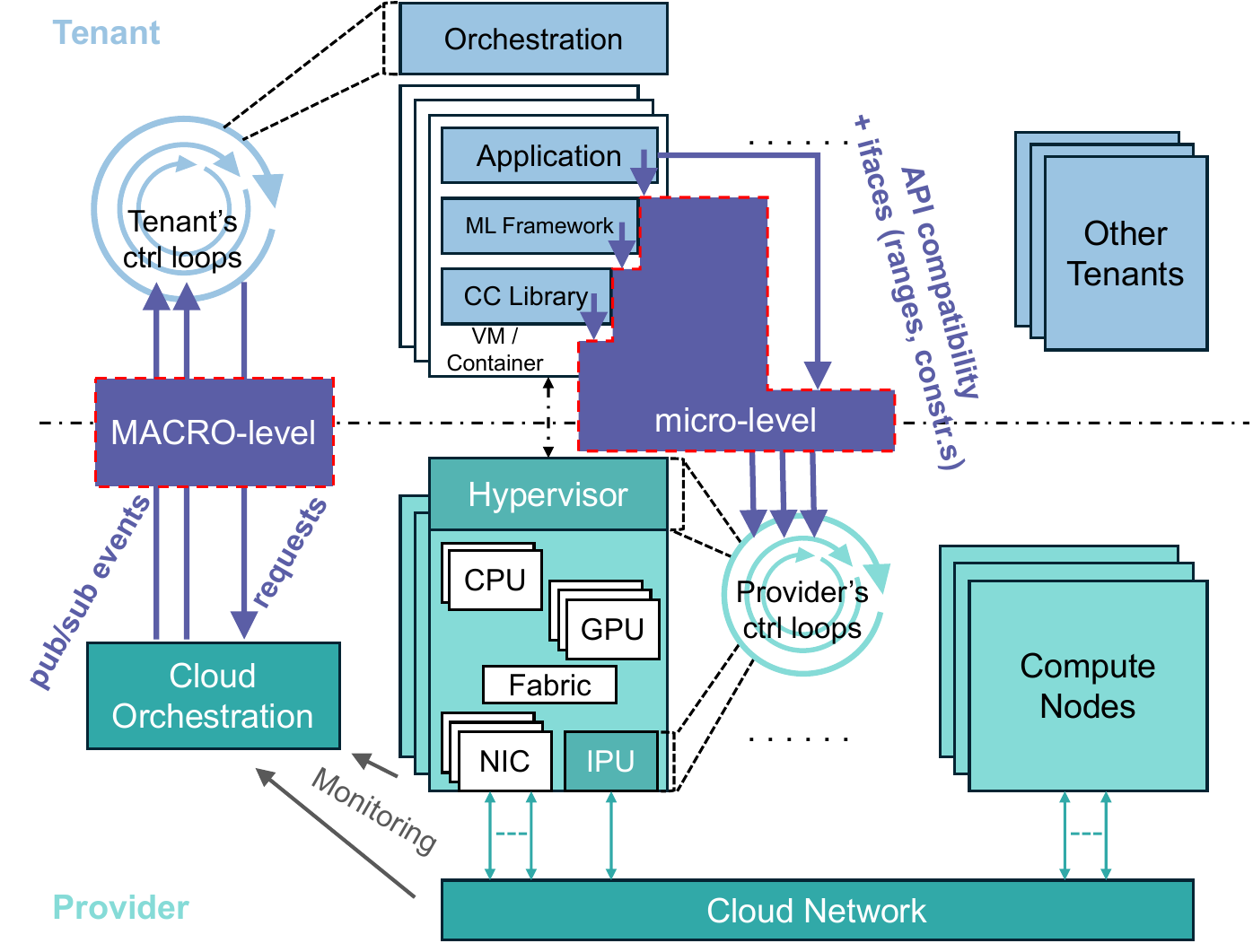}
    \caption{Architecture of \textcolor[HTML]{5B5EA6}{\AIsolate} and overview of interactions between \textcolor[HTML]{9BC4E2}{tenant} and \textcolor[HTML]{54B4B2}{provider} to realize cross-layer optimization opportunities via \mybox{\textcolor{white}{micro-}} and \mybox{\textcolor{white}{MACRO-}}level cloud abstractions. \small{\normalfont Dashed call-outs indicate potential realization loci (e.g., at IPUs~\cite{intel-ipu}) of optimized control loops.}}
    \label{fig:aisolate}
\end{figure}

To address this gap, we propose widening traditional cloud abstractions to enable tenant-provider collaboration. Specifically, these abstractions expand cloud interface boundaries to provide tenants with additional information typically unavailable to them and enable providers to optimize workloads more effectively.
\cref{fig:aisolate} outlines the architecture of our proposal, \AIsolate, and illustrates interactions between tenants and providers to enable cross-layer optimization opportunities (\cref{sec:opportunities}) via \emph{micro-} and \emph{macro-}level abstractions (\cref{sec:aisolate}). Compared to the status quo, \AIsolate introduces two key innovations: (1) tenants' control loops are informed by infrastructure-level insights (e.g., a shift in network load), and (2) providers' control loops are guided by workload-specific requirements (e.g., required communication patterns), integrating optimizations that tenants could only achieve with perfect infrastructure visibility in an ideal scenario.

\definecolor{AIcolor}{HTML}{5B5EA6}

\newtcbox{\mybox}[1][AIcolor]{on line,size=small,
arc=0pt,outer arc=0pt,colback=#1,colframe=red,
boxsep=0pt,left=1pt,right=1pt,top=2pt,bottom=2pt,
boxrule=0pt,bottomrule=1pt,toprule=1pt}

Recent work has introduced ad hoc approaches for optimizing AI workloads in specific cases~\cite{wu2024mccs,mai2020kungfu,duan2024parcae}. In contrast, \AIsolate advocates a general framework for cooperative optimization. Workload Intelligence (WI)~\cite{huang2024workloadintelligence} proposes dynamic, bi-directional communication for generic cloud cooperation. While \AIsolate shares some goals with WI, it departs from preserving traditional VM abstractions and VM-oriented mechanisms. Instead, it elevates its abstractions closer to the workload and refactors functionality to leverage unique workloads characteristics, such as their accelerator-driven, iterative nature, performance predictability, proximity to the network, and tight synchronization.

This paper aims to spark discussion in the systems research community on rethinking cloud abstractions tailored to AI workloads and their potential benefits. We highlight cross-layer optimization opportunities these abstractions could unlock. We also explore the challenges and trade-offs, such as standardization, privacy, and collaboration incentives, and propose a research agenda to bridge the tenant-provider gap, fostering efficient, sustainable, and scalable AI workloads in cloud environments.

\section{Today’s Cloud Fails AI Workloads}
\label{sec:background}

Tenants today can choose from a range of cloud-based options to run AI
workloads, including managed Machine Learning as a Service (MLaaS)
platforms, Infrastructure as a Service (IaaS), or hybrid
configurations. Each model presents trade-offs between abstraction and
control, and current systems fall short in enabling effective
cross-layer optimization.

\textbf{MLaaS platforms} (e.g., Google Vertex
AI~\cite{google-vertexai}, AWS SageMaker~\cite{aws-sagemaker}) offer
convenient interfaces for launching and scaling training
jobs. However, they provide little to no insight into low-level
resource behavior or cluster conditions. For example, systems like
Google's Reduction Server~\cite{google-reductionserver} optimize
communication, but tenants cannot influence or adapt those strategies
at runtime. Furthermore, AutoML features limit control over model
architecture and hyperparameters, while proprietary APIs restrict
portability across clouds.

\textbf{IaaS-based deployments}, in contrast, grant tenants full
control over infrastructure via virtual machines, Kubernetes clusters,
and direct access to GPUs or TPUs. However, this control comes at the
cost of responsibility. Tenants must manually manage aspects such as
job placement, communication library selection, and failure
recovery---tasks that are both complex and decoupled from provider-level
information.

These limitations lead to three pervasive problems:

\begin{compactitem}[\labelitemi]
\item \textbf{Jobs are static when they should be adaptive:} Most training
  jobs are configured statically, with fixed resource allocations and
  learning settings. This leads to inefficiencies when network
  conditions, cluster load, or hardware performance fluctuates
  mid-run. For instance, systems like KungFu~\cite{mai2020kungfu} and
  Parcae~\cite{duan2024parcae} show that adapting batch sizes or
  optimizer settings dynamically can improve convergence and
  throughput. Today, such adaptations require extensive custom
  infrastructure.

\item \textbf{Tenants and providers operate in the dark:} Tenants
  choose resources blindly, without visibility into current cluster
  utilization or topology. Communication libraries like
  NCCL~\cite{nccl} select algorithms (e.g., ring or tree) at
  initialization, based on assumptions that may no longer hold as the
  job progresses. Providers, on the other hand, lack hooks to
  influence job behavior in response to failures or congestion. This
  gap prevents runtime adaptation or efficient network-aware
  execution~\cite{wu2024mccs,cao2024crux,liu2024teccl}.

\item \textbf{Resource allocation is clumsy and manual:} Cluster
  schedulers assign resources based on static tenant requests, even
  though ideal allocations vary over time. Advanced scheduling systems
  like Pollux~\cite{qiao2021pollux} and
  Gandiva~\cite{chaudhary2020gandivafair} demonstrate the benefits of
  adaptive scaling, but require full control over both workload and
  infrastructure. Public clouds cannot offer this level of control
  without proper abstractions.
\end{compactitem}

\section{Coordinating AI Workloads with \AIsolate}
\label{sec:aisolate}

The lack of abstractions to enable cooperative optimization between tenants and providers significantly hinders the efficient execution of AI workloads in cloud environments. To address this challenge, we propose \AIsolate, a set of abstractions designed to foster cooperative optimization between tenants and providers.

Broadly, AI workloads can be understood through two distinct timescales of control: \textit{micro} and \textit{macro}. This distinction highlights the varying nature of optimizations at each level and helps clarify the roles of tenants and providers in executing control loops and sharing information (c.f. \cref{fig:aisolate}).

\smartparagraph{Micro-level control:}  
The micro level focuses on fine-grained control of operations within a tenant's workload. These operations typically occur within a single training iteration or an inference request. While workloads at this level exhibit predictable patterns across repeated cycles, their performance can still be influenced by the infrastructure and its dynamics.  

Tenants have visibility into key workload-specific aspects such as communication patterns, task priorities, and fine-grained performance-cost trade-offs. However, the timeliness of control loops at the micro level is critical, as adaptation decisions require up-to-date infrastructure-level information. Consequently, we propose that providers should execute the control loop at the micro level, leveraging requirements specified by tenants. With their detailed knowledge of the underlying infrastructure, providers can make decisions such as selecting optimal communication algorithms, adjusting batch sizes or compression ratios to meet latency targets, or determining the best points for checkpointing the model.

\smartparagraph{Macro-level control:}  
The macro level focuses on workload adaptation over coarser timescales, such as training epochs, multiple request batches, or in response to major trigger events (e.g., failures or resource availability changes). Examples of macro-level optimizations include scaling the number of workers, modifying parallelization strategies, or adjusting hyperparameters to improve training or inference efficiency.  

These strategic optimizations require a broader view of workload objectives and system resources. Tenants are best positioned to execute control loops at the macro level, as they have detailed insights into workload requirements, such as expected training duration or desired accuracy targets. To enable effective macro-level decisions, providers should supply infrastructure-level updates, such as resource availability changes or failure information. Importantly, these updates may be provided in aggregate or coarse-grained forms to safeguard sensitive infrastructure details.

\vspace{0.1em}
By distinguishing between micro- and macro-level control and assigning responsibility accordingly, \AIsolate enables a balanced approach to cross-layer optimization. Providers can leverage their expertise for real-time, fine-grained adaptations, while tenants retain control over strategic, workload-level decisions, creating a cooperative framework for optimizing AI workloads in multi-tenant cloud environments.

Given this high-level structuring of \AIsolate, the next critical questions to address are: (1) what abstractions are necessary to enable these cooperative control loops, and (2) how can they be realized in practice while minimizing adoption overhead? We now present preliminary approaches to tackle these challenges.

\begin{table}[t]
\resizebox{\columnwidth}{!}{%
\begin{tabular}{|l|p{4cm}|p{4.5cm}|l|}
\hline
\textbf{Opt.}       & \textbf{Example benefits}                                    & \textbf{Cross-layer approach}                                           & \textbf{Scope}       \\
\hline
O1 & $\downarrow$ 20\% costs \cite{mai2020kungfu}                                  & Joint hyperparameter tuning and resource allocation & Train            \\
\hline
O2 & $\uparrow$ 2.4$\times$ comm. perf. \cite{wu2024mccs}                            & Algorithm selection, collectives-as-a-service   & Train, Inf \\
\hline
O3 & $\downarrow$ JCT by 37-50\% \cite{qiao2021pollux}                              & Resource allocation                             & Train            \\
\hline
O4 & $\uparrow$ 3.5$\times$ speedup \cite{zheng2022alpa}                           & Constraint-guided partitioning                  & Train, Inf \\
\hline
O5 & $\uparrow$ 1.9$\times$ efficiency \cite{he2024unicron}                           & Managed checkpointing                           & Train            \\
\hline

O6 & $\uparrow$ 22.61$\times$ up time between failures \cite{xiong2024superbench} & Joint collection of observability data                 & Train \\
\hline
O7 & $\downarrow$ 53\% energy, 38\% operational carbon emissions \cite{stojkovic2024dynamollm} & Joint device-workload tuning                    & Train, Inf \\
\hline
\end{tabular}
}
\caption{Benefits of optimizations in literature and potential cross-layer approaches (\cref{sec:opportunities}), with primary workload scope.}
\label{tab:examples}
\end{table}
\subsection{Redefined cloud abstractions}

We outline key principles and requirements these abstractions must satisfy. Following the division into micro- and macro-level control, we structure the design considerations for these abstractions accordingly.

\smartparagraph{Micro-level abstractions:}
Micro-level abstractions should prioritize well-defined APIs that align with existing AI workloads, enabling \AIsolate to be adopted as a drop-in module with minimal code changes. These abstractions must support API compatibility while introducing additional interfaces for tenants to specify requirements and preferences (buy-in). Typically, these interfaces would allow tenants to define ranges or constraints to guide the provider's  decisions.

For instance, consider collective communication, where tenants often rely on popular libraries like NCCL. These libraries, however, lack adaptability and network awareness, which can result in suboptimal algorithm selection and degraded performance~\cite{wu2024mccs}. An abstraction for collective communication in \AIsolate could offer a compatible API that tenants already use while delegating underlying decisions -- such as algorithm selection -- and data path operations to providers. Providers, in turn, could leverage real-time network conditions to optimize performance dynamically.
A similar approach has demonstrated promise in prior works~\cite{wu2024mccs,chen2022nethint}, which leverage network-aware abstractions to enhance communication performance in distributed ML workloads. The main challenge lies in ensuring these abstractions remain transparent to tenants, requiring minimal changes to existing codes while empowering providers to perform infrastructure-level optimizations effectively. Thus, micro-level abstractions must be lightweight, non-intrusive, and compatible with existing AI frameworks and libraries.

Another example is model checkpointing, a critical component of large-scale model training~\cite{lian2024ucp,wang2024fastpersist,maurya2024datastatesllm}. UCP~\cite{lian2024ucp} demonstrates an API that enables efficient and flexible checkpointing across a range of parallelization strategies. With \AIsolate, tenants could use abstractions to specify desired checkpointing frequencies, the importance of individual checkpoints, and trade-offs between overhead and fault tolerance. Providers could then optimize checkpointing~\cite{wang2023gemini,gupta2024jitcheckpointing} based on tenant preferences and real-time infrastructure conditions, while ensuring compatibility with UCP's data format.

\smartparagraph{Macro-level abstractions:}
While maximizing API compatibility is critical for micro-level abstractions, macro-level abstractions allow for a more flexible, clean-slate approach to interface design.
The primary goal of macro-level abstractions is to provide tenants with an interface that is both expressive enough to convey meaningful information and narrow enough to protect sensitive infrastructure details. One potential design is for providers to expose a pub/sub interface, enabling tenants to subscribe to infrastructure-level events and updates. For example, if a provider frees up a GPU, it could send a notification to all subscribed tenants. To prevent contention when multiple tenants attempt to allocate the same resource, the system could implement mechanisms such as priority-based allocation, reservation windows, or price-based auctioning~\cite{chaudhary2020gandivafair}. To ensure broad applicability, this interface should be standardized across providers.

Importantly, tenants are not restricted to reacting solely to provider-generated events. They can also initiate adaptations based on workload conditions, such as training progress, accuracy milestones, or monitoring metrics, as per current practice~\cite{mai2020kungfu}. \AIsolate's abstractions simply enhance their decision-making capabilities by providing richer infrastructure information, to adapt workloads more strategically.

Given the coarser nature of macro-level adaptations, some scenarios may require tenants and providers to engage several rounds of negotiation to align on strategic decisions, as in \cite{chaudhary2020gandivafair,hindman2011mesos}. This process could involve tenants specifying high-level goals, constraints, and preferences, while providers suggest optimizations that address these requirements.

\subsection{Control loops}

Control loops are not a novel concept introduced by \AIsolate; both tenants and providers already execute control loops independently. However, these loops often operate in isolation, limiting their potential for optimization.
Instead, the key contribution of \AIsolate lies in restructuring existing control logic and introducing interfaces that enable new opportunities for cooperation. We now elaborate on how tenants and providers can execute control loops cooperatively, sketching a plausible realization of \AIsolate's abstractions.

\smartparagraph{Provider side:}  
To meet the requirement for fast, data-path-like execution, the provider's control loops could be implemented within the worker's hypervisor or a provider-managed container. A proven reference model for such an approach is Google's Andromeda~\cite{dalton2018andromeda}. Additionally, recent advancements in Infrastructure Processing Units (IPUs)~\cite{intel-ipu} open new possibilities. IPUs are isolated compute units that operate as a separate trusted domain. This isolation allows providers to execute control loops securely and efficiently, offering a fast path for infrastructure-level optimizations while maintaining isolation from tenants.
In either case, we must avoid exposing sensitive infrastructure details to tenants.

In restructuring control loops, some burden may shift to providers, as certain data-path routines previously executed by tenants may now be handled by the provider. For instance, in the case of collective communication, the provider becomes responsible for algorithm selection, as discussed earlier. But simply executing algorithm selection within the provider's control loop and feeding the results back to the tenant would be both cumbersome and potentially leaky. Instead, as demonstrated by \cite{wu2024mccs}, the provider initiates collective communication operations directly through the API compatibility layer envisioned in \AIsolate. Encouragingly, offloading collectives to IPUs (or DPUs) has already shown significant performance benefits in existing systems~\cite{gu2024omniccl,bayatpour2021bluesmpi,graham2024bluefield}.

\smartparagraph{Tenant side:}  
Control loops executed by the tenant retain their existing flexibility and remain within the tenant's own orchestration layer. \AIsolate does not impose restrictions on how tenants manage their workloads. Instead, it enhances the tenant's control logic by enabling them to subscribe to and act upon relevant provider-supplied events. This additional context enriches the tenant's decision-making process, allowing infrastructure updates into their optimizations.

\section{Cross-Layer Benefits of \AIsolate}
\label{sec:opportunities}

We now illustrate how \AIsolate enables a range of cross-layer optimizations that are difficult or impossible to realize in today’s cloud environments. In each case, \AIsolate acts as the missing link
between tenant-level logic and provider-level infrastructure, enabling
both sides to make more informed, coordinated decisions.

\smartparagraph{O1-Runtime adaptation:}
In cloud environments, system conditions and workload requirements evolve continuously. Traditional AI workloads, particularly training, operate with fixed configurations set at initialization, making them inflexible to changes in resource availability or workload dynamics during execution. Approaches such as those described in \cite{mai2020kungfu,duan2024parcae} enable dynamic adjustments to key training parameters (e.g., learning rate, batch size, gradient accumulation) and system configurations (e.g., communication protocols, resource usage). This allows tenants to respond to shifts in cluster load, network congestion, or performance variability, optimizing training efficiency and convergence speed.

\smartparagraph{O2-Collective communication:}
Collective communication libraries suffer from inefficiencies due to a lack of network awareness and adaptability~\cite{wu2024mccs,cao2024crux,liu2024teccl}. For instance, selecting the best communication algorithm (e.g., ring- vs. tree-based) relies on knowledge of network topology and utilization, which tenants cannot access. Consequently, suboptimal choices are common, and libraries such as NCCL~\cite{nccl} that decide on strategies at initialization may make choices that become less effective as conditions change. Additionally, common strategies often rely on assumptions about network paths that may not align with the actual physical configuration, resulting in unexploited optimization potential.

\smartparagraph{O3-Cluster scheduling:}
Another issue lies with scheduling, where tenants must configure resources manually. Suboptimal and static configurations can lead to queuing delays or inefficient resource usage. The optimal resource allocation varies dynamically~\cite{qiao2021pollux,chaudhary2020gandivafair}, yet tenants lack adaptive scheduling tools that could adjust resource allocation based on real-time cluster load, or resource availability.

\smartparagraph{O4-Parallelization and partitioning:}
For large models, effective distributed execution depends on complex combinations of data, operator, and pipeline parallelism~\cite{shoeybi2020megatronlm,hsia2024madmax}. While automation of parallelization through planning tools could greatly accelerate model execution, these tools expect access to the underlying cluster topology~\cite{tarnawski2021piper,li2022amp,wang2019tofu,cai2022tensoropt,zheng2022alpa,jia2022whale,lin2024nnscaler}. Planning optimal partitions also depends on dynamic conditions like workload composition, network load, and resource availability, requiring tenant monitoring combined with provider-accessible infrastructure-level information.

\smartparagraph{O5-Reliability:}
Resilience in large-scale training is a critical challenge~\cite{reka-llms-wilderness}. Failures cause delays and often require manual intervention, as tenants lack infrastructure visibility. Tenant-led systems, relying on redundancy, can incur unsustainable overheads~\cite{he2024unicron}. This highlights the need for efficient checkpointing~\cite{wang2023gemini,wang2024fastpersist,cai2024mocsystem,maurya2024datastatesllm,gupta2024jitcheckpointing}, which could benefit from deeper infrastructure visibility. Providers are well-positioned to manage fault tolerance holistically, avoiding the inefficiencies of treating training tasks as isolated black-boxes.

\smartparagraph{O6-Availability:} Recovery techniques, e.g., check pointing, mitigate the impact of failures, but such failures still cause a significant loss in revenue, e.g., as much as a \$30K~\cite{qian2024alibabahpn} .  Recent work~\cite{xiong2024superbench} shows that the mean time between failures can be increased by 22.61×\% if defective components are detected proactively and removed quickly. However, such approaches require knowledge of a tenant's workloads and events coupled with the provider's ability to perform proactive measurements using information about their infrastructure. Tenant and provider cooperation ensures that diagnostics framework have access to the application level events and signals needed to disambiguate problems and the infrastructure level observability data needed to triage and identify the root cause of these problems.

\smartparagraph{O7-Energy management:}
Energy consumption is a critical concern for AI workloads~\cite{wu2024sustainableai}. Achieving efficiency requires aligning workload characteristics with infrastructure capabilities~\cite{chung2024perseus,you2023zeus}. For example, dynamic inference cluster reconfiguration can optimize energy and cost while meeting performance SLOs~\cite{stojkovic2024dynamollm}. However, such optimizations demand insights into workload needs and infrastructure (e.g., assessing carbon intensity), which are often siloed. Cooperative optimization bridges this gap, enabling better strategies.

\vspace{0.1em}
Recent works (\cref{tab:examples}) show gains in these areas (O1-7), with cross-layer cooperation expected to amplify the benefits.
Beyond these examples, other potential areas for tenant-provider cooperation include profiling and troubleshooting tools, adaptive workload management~\cite{patke2024qlm}, and adaptive communication compression~\cite{abdelmoniem2021dc2,xin2023kimad,markov2024lgreco,chen2024packettrimming}.

\section{Discussion}

\smartparagraph{Adoption.}
The adoption of \AIsolate is underpinned by a natural alignment of incentives between tenants and providers. Tenants gain improved performance, efficiency, availability and resilience for their workloads, while providers benefit from better infrastructure utilization and the ability to offer differentiated, value-added services. 
Moreover, a buy-in approach affords incremental deployability. Tenants can adopt \AIsolate's abstractions at their own pace, without disrupting existing workloads. Providers, in turn, can continue handling workloads from non-adopting tenants using current practices while leveraging the additional insights provided by \AIsolate tenants. This gradual deployment model ensures a smooth transition toward cooperative optimization.

\smartparagraph{Timing and enablers.}
The timing of \AIsolate's introduction is critical. If introduced too early, it risks stifling innovation by imposing abstractions on a still-evolving ecosystem. Conversely, if introduced too late, the accumulated technical debt -- an issue particularly pronounced in ML systems~\cite{sculley2015technicaldebt} -- and the complexity of standardizing entrenched systems may diminish its impact.  
In recent years, the pace of innovation in AI and workload deployment has been relentless. However, the ecosystem has begun to coalesce around pre-canned recipes and best practices, with systems and libraries like vLLM~\cite{kwon2023vllm}, DeepSpeed~\cite{rajbhandari2020zero}, Megatron-LM~\cite{shoeybi2020megatronlm}, NCCL, and Huggingface's Transformers~\cite{wolf2020transformers} becoming de-facto standards. This growing stability signals that the ecosystem is now ripe for a new layer of abstraction, as we envision.

\smartparagraph{Why not generalize beyond AI?} 
One might critique that the issues \AIsolate addresses apply to general large-scale distributed systems, not just AI workloads. While true, these challenges are particularly acute for AI workloads due to their tightly synchronous nature and the high infrastructure costs associated with them.
Additionally, AI workloads are typically iterative, accelerator-driven, and less control-flow oriented, making them more predictable. This predictability suggests that the optimization problem is more tractable compared to general distributed systems~\cite{huang2024workloadintelligence}, offering substantial potential gains. For AI, even marginal improvements can translate into significant cost savings and performance enhancements, justifying a focused approach.

Attempting to generalize beyond AI risks overwhelming complexity. Crafting one-size-fits-all solutions while maintaining narrow, meaningful interfaces for cooperative optimization is a significant challenge~\cite{huang2024workloadintelligence}.
To achieve practicality, prior work has often narrowed its focus, as seen in collectives~\cite{wu2024mccs}, network awareness in DCs~\cite{chen2022nethint} and P2P applications~\cite{xie2008p4p}, or in domains like virtualized congestion control~\cite{cronkite2016virtualized,he2016acdc}, ISP-CDN-content provider cooperation~\cite{frank2013cdnisp,jiang2014eona}, and serverless-cloud cooperation~\cite{zhao2024serverless,bilal2023serverless}. Our work applies a similar philosophy in the AI workloads domain.

\smartparagraph{Could tenants do it all themselves?}
Some argue that tenants, particularly those operating thousands of GPUs in the cloud, already experience a ``bare metal''-like environment with access to provider-specific interfaces for managing placement, network locality, and maintenance. While this may hold true for the largest cloud customers, these interfaces are often proprietary, non-standardized, and inaccessible to the broader tenant base. \AIsolate aims to democratize access to advanced optimizations by providing a standardized framework, making such capabilities available to a wider range of tenants and use cases. This democratization is a necessary step toward a shared responsibility for better sustainability of AI workloads~\cite{wu2024sustainableai}.

\smartparagraph{Immediate roadmap for \AIsolate.}
Beyond the discussion this paper aims to inspire, the immediate roadmap for \AIsolate is grounded in several foundational building blocks already produced by the research community.
A first proof-of-concept prototype for \AIsolate can leverage recently open-sourced tools and frameworks. These include tools for simulating AI job schedulers~\cite{agarwal2024blox}, emulating the behavior of distributed jobs~\cite{liu2024neuronabox}, and systems realizing collective communication as a service~\cite{wu2024mccs} and runtime adaptations~\cite{mai2020kungfu}. By integrating these components, an initial prototype could demonstrate the feasibility of \AIsolate's cooperative optimization vision and showcase the tangible benefits of cross-layer control loops for AI workloads.

Early efforts could then focus on the opportunities identified in \cref{sec:opportunities} and establish a framework for evaluating the impact of \AIsolate's abstractions on representative AI workloads. Such a framework would enable systematic assessment of the performance, efficiency, availability and resilience gains achievable through cooperative optimization.
Using published results, \cref{tab:examples} illustrates select benefits that are representative of cooperative optimizations that we expect from \AIsolate.

Several open avenues for exploration remain. Beyond the identified opportunities, further research could uncover additional potential for cooperative optimization. Another intriguing direction is the automation of control loop logic based on high-level intents, allowing tenants to specify their goals without needing to define intricate control mechanisms. Finally, addressing the privacy implications of sharing infrastructure information with tenants is a concern. \AIsolate abstractions must balance the need for transparency with the imperative to safeguard sensitive provider details.

By fostering cooperative optimization, \AIsolate can create a virtuous cycle of innovation and efficiency in multi-tenant cloud environments.

\begin{acks}
We are grateful to Amin Vahdat for his insightful feedback and suggestions. This work has been partially supported by Knut and Alice Wallenberg Foundation (Wallenberg Scholar Grant for Prof. Dejan Kosti\'c), as well as NSF grant 2212193.
\end{acks}

\balance

\bibliographystyle{ACM-Reference-Format}
\bibliography{main}


\begin{thebibliography}{62}


\ifx \showCODEN    \undefined \def \showCODEN     #1{\unskip}     \fi
\ifx \showDOI      \undefined \def \showDOI       #1{#1}\fi
\ifx \showISBNx    \undefined \def \showISBNx     #1{\unskip}     \fi
\ifx \showISBNxiii \undefined \def \showISBNxiii  #1{\unskip}     \fi
\ifx \showISSN     \undefined \def \showISSN      #1{\unskip}     \fi
\ifx \showLCCN     \undefined \def \showLCCN      #1{\unskip}     \fi
\ifx \shownote     \undefined \def \shownote      #1{#1}          \fi
\ifx \showarticletitle \undefined \def \showarticletitle #1{#1}   \fi
\ifx \showURL      \undefined \def \showURL       {\relax}        \fi
\providecommand\bibfield[2]{#2}
\providecommand\bibinfo[2]{#2}
\providecommand\natexlab[1]{#1}
\providecommand\showeprint[2][]{arXiv:#2}

\bibitem[Abdelmoniem and Canini(2021)]%
        {abdelmoniem2021dc2}
\bibfield{author}{\bibinfo{person}{Ahmed~M. Abdelmoniem} {and}
  \bibinfo{person}{Marco Canini}.} \bibinfo{year}{2021}\natexlab{}.
\newblock
  \showarticletitle{\href{https://doi.org/10.1109/INFOCOM42981.2021.9488810}{{DC2:
  Delay-aware Compression Control for Distributed Machine Learning}}}. In
  \bibinfo{booktitle}{\emph{INFOCOM}}.
\newblock


\bibitem[Agarwal et~al\mbox{.}(2024)]%
        {agarwal2024blox}
\bibfield{author}{\bibinfo{person}{Saurabh Agarwal}, \bibinfo{person}{Amar
  Phanishayee}, {and} \bibinfo{person}{Shivaram Venkataraman}.}
  \bibinfo{year}{2024}\natexlab{}.
\newblock
  \showarticletitle{\href{https://doi.org/10.1145/3627703.3629583}{{Blox: A
  Modular Toolkit for Deep Learning Schedulers}}}. In
  \bibinfo{booktitle}{\emph{EuroSys}}.
\newblock


\bibitem[{Amazon Web Services}(2024)]%
        {aws-sagemaker}
\bibfield{author}{\bibinfo{person}{{Amazon Web Services}}.}
  \bibinfo{year}{2024}\natexlab{}.
\newblock \bibinfo{title}{AWS SageMaker}.
\newblock
\newblock
\urldef\tempurl%
\url{https://aws.amazon.com/sagemaker}
\showURL{%
\tempurl}


\bibitem[Bayatpour et~al\mbox{.}(2021)]%
        {bayatpour2021bluesmpi}
\bibfield{author}{\bibinfo{person}{Mohammadreza Bayatpour},
  \bibinfo{person}{Nick Sarkauskas}, \bibinfo{person}{Hari Subramoni},
  \bibinfo{person}{Jahanzeb Maqbool~Hashmi}, {and}
  \bibinfo{person}{Dhabaleswar~K. Panda}.} \bibinfo{year}{2021}\natexlab{}.
\newblock
  \showarticletitle{\href{https://doi.org/10.1007/978-3-030-78713-4_2}{{BluesMPI:
  Efficient MPI Non-blocking Alltoall Offloading Designs on Modern BlueField
  Smart NICs}}}. In \bibinfo{booktitle}{\emph{High Performance Computing}}.
\newblock


\bibitem[Bilal et~al\mbox{.}(2023)]%
        {bilal2023serverless}
\bibfield{author}{\bibinfo{person}{Muhammad Bilal}, \bibinfo{person}{Marco
  Canini}, \bibinfo{person}{Rodrigo Fonseca}, {and} \bibinfo{person}{Rodrigo
  Rodrigues}.} \bibinfo{year}{2023}\natexlab{}.
\newblock
  \showarticletitle{\href{https://doi.org/10.1145/3552326.3567506}{{With Great
  Freedom Comes Great Opportunity: Rethinking Resource Allocation for
  Serverless Functions}}}. In \bibinfo{booktitle}{\emph{EuroSys}}.
\newblock


\bibitem[Cai et~al\mbox{.}(2025)]%
        {cai2024mocsystem}
\bibfield{author}{\bibinfo{person}{Weilin Cai}, \bibinfo{person}{Le Qin}, {and}
  \bibinfo{person}{Jiayi Huang}.} \bibinfo{year}{2025}\natexlab{}.
\newblock
  \showarticletitle{\href{https://doi.org/10.1145/3676641.3716006}{{MoC-System:
  Efficient Fault Tolerance for Sparse Mixture-of-Experts Model Training}}}. In
  \bibinfo{booktitle}{\emph{ASPLOS}}.
\newblock


\bibitem[Cai et~al\mbox{.}(2022)]%
        {cai2022tensoropt}
\bibfield{author}{\bibinfo{person}{Zhenkun Cai}, \bibinfo{person}{Xiao Yan},
  \bibinfo{person}{Kaihao Ma}, \bibinfo{person}{Yidi Wu},
  \bibinfo{person}{Yuzhen Huang}, \bibinfo{person}{James Cheng},
  \bibinfo{person}{Teng Su}, {and} \bibinfo{person}{Fan Yu}.}
  \bibinfo{year}{2022}\natexlab{}.
\newblock
  \showarticletitle{\href{https://doi.org/10.1109/TPDS.2021.3132413}{{TensorOpt:
  Exploring the Tradeoffs in Distributed DNN Training With Auto-Parallelism}}}.
\newblock \bibinfo{journal}{\emph{IEEE Transactions on Parallel and Distributed
  Systems}} \bibinfo{volume}{33}, \bibinfo{number}{8} (\bibinfo{year}{2022}).
\newblock


\bibitem[Cao et~al\mbox{.}(2024)]%
        {cao2024crux}
\bibfield{author}{\bibinfo{person}{Jiamin Cao}, \bibinfo{person}{Yu Guan},
  \bibinfo{person}{Kun Qian}, \bibinfo{person}{Jiaqi Gao},
  \bibinfo{person}{Wencong Xiao}, \bibinfo{person}{Jianbo Dong},
  \bibinfo{person}{Binzhang Fu}, \bibinfo{person}{Dennis Cai}, {and}
  \bibinfo{person}{Ennan Zhai}.} \bibinfo{year}{2024}\natexlab{}.
\newblock
  \showarticletitle{\href{https://doi.org/10.1145/3651890.3672239}{{Crux:
  GPU-Efficient Communication Scheduling for Deep Learning Training}}}. In
  \bibinfo{booktitle}{\emph{SIGCOMM}}.
\newblock


\bibitem[Chaudhary et~al\mbox{.}(2020)]%
        {chaudhary2020gandivafair}
\bibfield{author}{\bibinfo{person}{Shubham Chaudhary},
  \bibinfo{person}{Ramachandran Ramjee}, \bibinfo{person}{Muthian Sivathanu},
  \bibinfo{person}{Nipun Kwatra}, {and} \bibinfo{person}{Srinidhi Viswanatha}.}
  \bibinfo{year}{2020}\natexlab{}.
\newblock
  \showarticletitle{\href{https://doi.org/10.1145/3342195.3387555}{{Balancing
  Efficiency and Fairness in Heterogeneous GPU Clusters for Deep Learning}}}.
  In \bibinfo{booktitle}{\emph{EuroSys}}.
\newblock


\bibitem[Chen et~al\mbox{.}(2022)]%
        {chen2022nethint}
\bibfield{author}{\bibinfo{person}{Jingrong Chen}, \bibinfo{person}{Hong
  Zhang}, \bibinfo{person}{Wei Zhang}, \bibinfo{person}{Liang Luo},
  \bibinfo{person}{Jeffrey Chase}, \bibinfo{person}{Ion Stoica}, {and}
  \bibinfo{person}{Danyang Zhuo}.} \bibinfo{year}{2022}\natexlab{}.
\newblock
  \showarticletitle{\href{https://www.usenix.org/conference/nsdi22/presentation/chen-jingrong}{{NetHint:
  White-Box Networking for Multi-Tenant Data Centers}}}. In
  \bibinfo{booktitle}{\emph{NSDI}}.
\newblock


\bibitem[Chen et~al\mbox{.}(2024)]%
        {chen2024packettrimming}
\bibfield{author}{\bibinfo{person}{Xiaoqi Chen}, \bibinfo{person}{Shay
  Vargaftik}, {and} \bibinfo{person}{Ran~Ben Basat}.}
  \bibinfo{year}{2024}\natexlab{}.
\newblock
  \showarticletitle{\href{https://doi.org/10.1145/3696348.3696880}{{When ML
  Training Cuts Through Congestion: Just-in-Time Gradient Compression via
  Packet Trimming}}}. In \bibinfo{booktitle}{\emph{HotNets}}.
\newblock


\bibitem[Chung et~al\mbox{.}(2024)]%
        {chung2024perseus}
\bibfield{author}{\bibinfo{person}{Jae-Won Chung}, \bibinfo{person}{Yile Gu},
  \bibinfo{person}{Insu Jang}, \bibinfo{person}{Luoxi Meng},
  \bibinfo{person}{Nikhil Bansal}, {and} \bibinfo{person}{Mosharaf Chowdhury}.}
  \bibinfo{year}{2024}\natexlab{}.
\newblock
  \showarticletitle{\href{https://doi.org/10.1145/3694715.3695970}{{Reducing
  Energy Bloat in Large Model Training}}}. In \bibinfo{booktitle}{\emph{SOSP}}.
\newblock


\bibitem[Cronkite-Ratcliff et~al\mbox{.}(2016)]%
        {cronkite2016virtualized}
\bibfield{author}{\bibinfo{person}{Bryce Cronkite-Ratcliff},
  \bibinfo{person}{Aran Bergman}, \bibinfo{person}{Shay Vargaftik},
  \bibinfo{person}{Madhusudhan Ravi}, \bibinfo{person}{Nick McKeown},
  \bibinfo{person}{Ittai Abraham}, {and} \bibinfo{person}{Isaac Keslassy}.}
  \bibinfo{year}{2016}\natexlab{}.
\newblock
  \showarticletitle{\href{https://doi.org/10.1145/2934872.2934889}{{Virtualized
  Congestion Control}}}. In \bibinfo{booktitle}{\emph{SIGCOMM}}.
\newblock


\bibitem[Dalton et~al\mbox{.}(2018)]%
        {dalton2018andromeda}
\bibfield{author}{\bibinfo{person}{Michael Dalton}, \bibinfo{person}{David
  Schultz}, \bibinfo{person}{Jacob Adriaens}, \bibinfo{person}{Ahsan Arefin},
  \bibinfo{person}{Anshuman Gupta}, \bibinfo{person}{Brian Fahs},
  \bibinfo{person}{Dima Rubinstein}, \bibinfo{person}{Enrique~Cauich Zermeno},
  \bibinfo{person}{Erik Rubow}, \bibinfo{person}{James~Alexander Docauer},
  \bibinfo{person}{Jesse Alpert}, \bibinfo{person}{Jing Ai},
  \bibinfo{person}{Jon Olson}, \bibinfo{person}{Kevin DeCabooter},
  \bibinfo{person}{Marc de Kruijf}, \bibinfo{person}{Nan Hua},
  \bibinfo{person}{Nathan Lewis}, \bibinfo{person}{Nikhil Kasinadhuni},
  \bibinfo{person}{Riccardo Crepaldi}, \bibinfo{person}{Srinivas Krishnan},
  \bibinfo{person}{Subbaiah Venkata}, \bibinfo{person}{Yossi Richter},
  \bibinfo{person}{Uday Naik}, {and} \bibinfo{person}{Amin Vahdat}.}
  \bibinfo{year}{2018}\natexlab{}.
\newblock
  \showarticletitle{\href{https://www.usenix.org/conference/nsdi18/presentation/dalton}{{Andromeda:
  Performance, Isolation, and Velocity at Scale in Cloud Network
  Virtualization}}}. In \bibinfo{booktitle}{\emph{NSDI}}.
\newblock


\bibitem[Duan et~al\mbox{.}(2024)]%
        {duan2024parcae}
\bibfield{author}{\bibinfo{person}{Jiangfei Duan}, \bibinfo{person}{Ziang
  Song}, \bibinfo{person}{Xupeng Miao}, \bibinfo{person}{Xiaoli Xi},
  \bibinfo{person}{Dahua Lin}, \bibinfo{person}{Harry Xu},
  \bibinfo{person}{Minjia Zhang}, {and} \bibinfo{person}{Zhihao Jia}.}
  \bibinfo{year}{2024}\natexlab{}.
\newblock
  \showarticletitle{\href{https://www.usenix.org/conference/nsdi24/presentation/duan}{{Parcae:
  Proactive, Liveput-Optimized DNN Training on Preemptible Instances}}}. In
  \bibinfo{booktitle}{\emph{NSDI}}.
\newblock


\bibitem[Frank et~al\mbox{.}(2013)]%
        {frank2013cdnisp}
\bibfield{author}{\bibinfo{person}{Benjamin Frank}, \bibinfo{person}{Ingmar
  Poese}, \bibinfo{person}{Yin Lin}, \bibinfo{person}{Georgios Smaragdakis},
  \bibinfo{person}{Anja Feldmann}, \bibinfo{person}{Bruce Maggs},
  \bibinfo{person}{Jannis Rake}, \bibinfo{person}{Steve Uhlig}, {and}
  \bibinfo{person}{Rick Weber}.} \bibinfo{year}{2013}\natexlab{}.
\newblock
  \showarticletitle{\href{https://doi.org/10.1145/2500098.2500103}{{Pushing
  CDN-ISP Collaboration to the Limit}}}.
\newblock \bibinfo{journal}{\emph{SIGCOMM Comput. Commun. Rev.}}
  \bibinfo{volume}{43}, \bibinfo{number}{3} (\bibinfo{year}{2013}).
\newblock


\bibitem[{Google Cloud}(2024a)]%
        {google-vertexai}
\bibfield{author}{\bibinfo{person}{{Google Cloud}}.}
  \bibinfo{year}{2024}\natexlab{a}.
\newblock \bibinfo{title}{Google Vertex AI}.
\newblock
\newblock
\urldef\tempurl%
\url{https://cloud.google.com/vertex-ai}
\showURL{%
\tempurl}


\bibitem[{Google Cloud}(2024b)]%
        {google-reductionserver}
\bibfield{author}{\bibinfo{person}{{Google Cloud}}.}
  \bibinfo{year}{2024}\natexlab{b}.
\newblock \bibinfo{title}{Optimize Training Performance with Reduction Server
  in Vertex AI}.
\newblock
\newblock
\urldef\tempurl%
\url{https://cloud.google.com/blog/topics/developers-practitioners/optimize-training-performance-reduction-server-vertex-ai}
\showURL{%
\tempurl}


\bibitem[Graham et~al\mbox{.}(2024)]%
        {graham2024bluefield}
\bibfield{author}{\bibinfo{person}{Richard Graham}, \bibinfo{person}{George
  Bosilca}, \bibinfo{person}{Yong Qin}, \bibinfo{person}{Bradley Settlemyer},
  \bibinfo{person}{Gilad Shainer}, \bibinfo{person}{Craig Stunkel},
  \bibinfo{person}{Geoffroy Vallee}, \bibinfo{person}{Brody Williams},
  \bibinfo{person}{Gerardo Cisneros-Stoianowski}, \bibinfo{person}{Sebastian
  Ohlmann}, {and} \bibinfo{person}{Markus Rampp}.}
  \bibinfo{year}{2024}\natexlab{}.
\newblock
  \showarticletitle{\href{https://doi.org/10.23919/ISC.2024.10528935}{{Optimizing
  Application Performance with BlueField: Accelerating Large-Message Blocking
  and Nonblocking Collective Operations}}}. In \bibinfo{booktitle}{\emph{ISC}}.
\newblock


\bibitem[Gu et~al\mbox{.}(2024)]%
        {gu2024omniccl}
\bibfield{author}{\bibinfo{person}{Tongzhou Gu}, \bibinfo{person}{Jiawei Fei},
  {and} \bibinfo{person}{Marco Canini}.} \bibinfo{year}{2024}\natexlab{}.
\newblock
  \showarticletitle{\href{https://doi.org/10.1145/3672198.3673804}{{OmNICCL:
  Zero-cost Sparse AllReduce with Direct Cache Access and SmartNICs}}}. In
  \bibinfo{booktitle}{\emph{NAIC}}.
\newblock


\bibitem[Gupta et~al\mbox{.}(2024)]%
        {gupta2024jitcheckpointing}
\bibfield{author}{\bibinfo{person}{Tanmaey Gupta}, \bibinfo{person}{Sanjeev
  Krishnan}, \bibinfo{person}{Rituraj Kumar}, \bibinfo{person}{Abhishek
  Vijeev}, \bibinfo{person}{Bhargav Gulavani}, \bibinfo{person}{Nipun Kwatra},
  \bibinfo{person}{Ramachandran Ramjee}, {and} \bibinfo{person}{Muthian
  Sivathanu}.} \bibinfo{year}{2024}\natexlab{}.
\newblock
  \showarticletitle{\href{https://doi.org/10.1145/3627703.3650085}{{Just-In-Time
  Checkpointing: Low Cost Error Recovery from Deep Learning Training
  Failures}}}. In \bibinfo{booktitle}{\emph{EuroSys}}.
\newblock


\bibitem[He et~al\mbox{.}(2016)]%
        {he2016acdc}
\bibfield{author}{\bibinfo{person}{Keqiang He}, \bibinfo{person}{Eric Rozner},
  \bibinfo{person}{Kanak Agarwal}, \bibinfo{person}{Yu~(Jason) Gu},
  \bibinfo{person}{Wes Felter}, \bibinfo{person}{John Carter}, {and}
  \bibinfo{person}{Aditya Akella}.} \bibinfo{year}{2016}\natexlab{}.
\newblock
  \showarticletitle{\href{https://doi.org/10.1145/2934872.2934903}{{AC/DC TCP:
  Virtual Congestion Control Enforcement for Datacenter Networks}}}. In
  \bibinfo{booktitle}{\emph{SIGCOMM}}.
\newblock


\bibitem[He et~al\mbox{.}(2024)]%
        {he2024unicron}
\bibfield{author}{\bibinfo{person}{Tao He}, \bibinfo{person}{Xue Li},
  \bibinfo{person}{Zhibin Wang}, \bibinfo{person}{Kun Qian},
  \bibinfo{person}{Jingbo Xu}, \bibinfo{person}{Wenyuan Yu}, {and}
  \bibinfo{person}{Jingren Zhou}.} \bibinfo{year}{2024}\natexlab{}.
\newblock \bibinfo{title}{\href{https://arxiv.org/abs/2401.00134}{{Unicron:
  Economizing Self-Healing LLM Training at Scale}}}.
\newblock
\newblock
\showeprint[arxiv]{2401.00134}~[cs.DC]


\bibitem[Hindman et~al\mbox{.}(2011)]%
        {hindman2011mesos}
\bibfield{author}{\bibinfo{person}{Benjamin Hindman}, \bibinfo{person}{Andy
  Konwinski}, \bibinfo{person}{Matei Zaharia}, \bibinfo{person}{Ali Ghodsi},
  \bibinfo{person}{Anthony~D. Joseph}, \bibinfo{person}{Randy Katz},
  \bibinfo{person}{Scott Shenker}, {and} \bibinfo{person}{Ion Stoica}.}
  \bibinfo{year}{2011}\natexlab{}.
\newblock
  \showarticletitle{\href{https://www.usenix.org/conference/nsdi11/mesos-platform-fine-grained-resource-sharing-data-center}{{Mesos:
  A Platform for Fine-Grained Resource Sharing in the Data Center}}}. In
  \bibinfo{booktitle}{\emph{NSDI}}.
\newblock


\bibitem[Hsia et~al\mbox{.}(2024)]%
        {hsia2024madmax}
\bibfield{author}{\bibinfo{person}{Samuel Hsia}, \bibinfo{person}{Alicia
  Golden}, \bibinfo{person}{Bilge Acun}, \bibinfo{person}{Newsha Ardalani},
  \bibinfo{person}{Zachary DeVito}, \bibinfo{person}{Gu-Yeon Wei},
  \bibinfo{person}{David Brooks}, {and} \bibinfo{person}{Carole-Jean Wu}.}
  \bibinfo{year}{2024}\natexlab{}.
\newblock
  \showarticletitle{\href{https://doi.org/10.1109/ISCA59077.2024.00064}{{MAD-Max
  Beyond Single-Node: Enabling Large Machine Learning Model Acceleration on
  Distributed Systems}}}. In \bibinfo{booktitle}{\emph{ISCA}}.
\newblock


\bibitem[Huang et~al\mbox{.}(2024)]%
        {huang2024workloadintelligence}
\bibfield{author}{\bibinfo{person}{Lexiang Huang}, \bibinfo{person}{Anjaly
  Parayil}, \bibinfo{person}{Jue Zhang}, \bibinfo{person}{Xiaoting Qin},
  \bibinfo{person}{Chetan Bansal}, \bibinfo{person}{Jovan Stojkovic},
  \bibinfo{person}{Pantea Zardoshti}, \bibinfo{person}{Pulkit Misra},
  \bibinfo{person}{Eli Cortez}, \bibinfo{person}{Raphael Ghelman},
  \bibinfo{person}{Íñigo Goiri}, \bibinfo{person}{Saravan Rajmohan},
  \bibinfo{person}{Jim Kleewein}, \bibinfo{person}{Rodrigo Fonseca},
  \bibinfo{person}{Timothy Zhu}, {and} \bibinfo{person}{Ricardo Bianchini}.}
  \bibinfo{year}{2024}\natexlab{}.
\newblock \bibinfo{title}{\href{https://arxiv.org/abs/2404.19143}{{Workload
  Intelligence: Punching Holes Through the Cloud Abstraction}}}.
\newblock
\newblock
\showeprint[arxiv]{2404.19143}~[cs.DC]


\bibitem[{Intel}(2023)]%
        {intel-ipu}
\bibfield{author}{\bibinfo{person}{{Intel}}.} \bibinfo{year}{2023}\natexlab{}.
\newblock \bibinfo{title}{Intel Infrastructure Processing Unit (Intel IPU) ASIC
  E2000}.
\newblock
\newblock
\urldef\tempurl%
\url{https://www.intel.com/content/www/us/en/products/details/network-io/ipu/e2000-asic.html}
\showURL{%
\tempurl}


\bibitem[Jia et~al\mbox{.}(2022)]%
        {jia2022whale}
\bibfield{author}{\bibinfo{person}{Xianyan Jia}, \bibinfo{person}{Le Jiang},
  \bibinfo{person}{Ang Wang}, \bibinfo{person}{Wencong Xiao},
  \bibinfo{person}{Ziji Shi}, \bibinfo{person}{Jie Zhang},
  \bibinfo{person}{Xinyuan Li}, \bibinfo{person}{Langshi Chen},
  \bibinfo{person}{Yong Li}, \bibinfo{person}{Zhen Zheng},
  \bibinfo{person}{Xiaoyong Liu}, {and} \bibinfo{person}{Wei Lin}.}
  \bibinfo{year}{2022}\natexlab{}.
\newblock
  \showarticletitle{\href{https://www.usenix.org/conference/atc22/presentation/jia-xianyan}{{Whale:
  Efficient Giant Model Training over Heterogeneous GPUs}}}. In
  \bibinfo{booktitle}{\emph{USENIX ATC}}.
\newblock


\bibitem[Jiang et~al\mbox{.}(2014)]%
        {jiang2014eona}
\bibfield{author}{\bibinfo{person}{Junchen Jiang}, \bibinfo{person}{Xi Liu},
  \bibinfo{person}{Vyas Sekar}, \bibinfo{person}{Ion Stoica}, {and}
  \bibinfo{person}{Hui Zhang}.} \bibinfo{year}{2014}\natexlab{}.
\newblock
  \showarticletitle{\href{https://doi.org/10.1145/2670518.2673878}{{EONA:
  Experience-Oriented Network Architecture}}}. In
  \bibinfo{booktitle}{\emph{HotNets}}.
\newblock


\bibitem[Jiang et~al\mbox{.}(2024)]%
        {jiang2024megascale}
\bibfield{author}{\bibinfo{person}{Ziheng Jiang}, \bibinfo{person}{Haibin Lin},
  \bibinfo{person}{Yinmin Zhong}, \bibinfo{person}{Qi Huang},
  \bibinfo{person}{Yangrui Chen}, \bibinfo{person}{Zhi Zhang},
  \bibinfo{person}{Yanghua Peng}, \bibinfo{person}{Xiang Li},
  \bibinfo{person}{Cong Xie}, \bibinfo{person}{Shibiao Nong},
  \bibinfo{person}{Yulu Jia}, \bibinfo{person}{Sun He},
  \bibinfo{person}{Hongmin Chen}, \bibinfo{person}{Zhihao Bai},
  \bibinfo{person}{Qi Hou}, \bibinfo{person}{Shipeng Yan},
  \bibinfo{person}{Ding Zhou}, \bibinfo{person}{Yiyao Sheng},
  \bibinfo{person}{Zhuo Jiang}, \bibinfo{person}{Haohan Xu},
  \bibinfo{person}{Haoran Wei}, \bibinfo{person}{Zhang Zhang},
  \bibinfo{person}{Pengfei Nie}, \bibinfo{person}{Leqi Zou},
  \bibinfo{person}{Sida Zhao}, \bibinfo{person}{Liang Xiang},
  \bibinfo{person}{Zherui Liu}, \bibinfo{person}{Zhe Li},
  \bibinfo{person}{Xiaoying Jia}, \bibinfo{person}{Jianxi Ye},
  \bibinfo{person}{Xin Jin}, {and} \bibinfo{person}{Xin Liu}.}
  \bibinfo{year}{2024}\natexlab{}.
\newblock
  \showarticletitle{\href{https://www.usenix.org/conference/nsdi24/presentation/jiang-ziheng}{{MegaScale:
  Scaling Large Language Model Training to More Than 10,000 GPUs}}}. In
  \bibinfo{booktitle}{\emph{NSDI}}.
\newblock


\bibitem[Kwon et~al\mbox{.}(2023)]%
        {kwon2023vllm}
\bibfield{author}{\bibinfo{person}{Woosuk Kwon}, \bibinfo{person}{Zhuohan Li},
  \bibinfo{person}{Siyuan Zhuang}, \bibinfo{person}{Ying Sheng},
  \bibinfo{person}{Lianmin Zheng}, \bibinfo{person}{Cody~Hao Yu},
  \bibinfo{person}{Joseph Gonzalez}, \bibinfo{person}{Hao Zhang}, {and}
  \bibinfo{person}{Ion Stoica}.} \bibinfo{year}{2023}\natexlab{}.
\newblock
  \showarticletitle{\href{https://doi.org/10.1145/3600006.3613165}{{Efficient
  Memory Management for Large Language Model Serving with PagedAttention}}}. In
  \bibinfo{booktitle}{\emph{SOSP}}.
\newblock


\bibitem[Li et~al\mbox{.}(2022)]%
        {li2022amp}
\bibfield{author}{\bibinfo{person}{Dacheng Li}, \bibinfo{person}{Hongyi Wang},
  \bibinfo{person}{Eric Xing}, {and} \bibinfo{person}{Hao Zhang}.}
  \bibinfo{year}{2022}\natexlab{}.
\newblock
  \showarticletitle{\href{https://proceedings.neurips.cc/paper_files/paper/2022/file/2b4bfa1cebe78d125fefd7ea6ffcfc6d-Paper-Conference.pdf}{{AMP:
  Automatically Finding Model Parallel Strategies with Heterogeneity
  Awareness}}}. In \bibinfo{booktitle}{\emph{NeurIPS}}.
\newblock


\bibitem[Lian et~al\mbox{.}(2024)]%
        {lian2024ucp}
\bibfield{author}{\bibinfo{person}{Xinyu Lian}, \bibinfo{person}{Sam~Ade
  Jacobs}, \bibinfo{person}{Lev Kurilenko}, \bibinfo{person}{Masahiro Tanaka},
  \bibinfo{person}{Stas Bekman}, \bibinfo{person}{Olatunji Ruwase}, {and}
  \bibinfo{person}{Minjia Zhang}.} \bibinfo{year}{2024}\natexlab{}.
\newblock \bibinfo{title}{\href{https://arxiv.org/abs/2406.18820}{{Universal
  Checkpointing: Efficient and Flexible Checkpointing for Large Scale
  Distributed Training}}}.
\newblock
\newblock
\showeprint[arxiv]{2406.18820}~[cs.DC]


\bibitem[Lin et~al\mbox{.}(2024)]%
        {lin2024nnscaler}
\bibfield{author}{\bibinfo{person}{Zhiqi Lin}, \bibinfo{person}{Youshan Miao},
  \bibinfo{person}{Quanlu Zhang}, \bibinfo{person}{Fan Yang},
  \bibinfo{person}{Yi Zhu}, \bibinfo{person}{Cheng Li}, \bibinfo{person}{Saeed
  Maleki}, \bibinfo{person}{Xu Cao}, \bibinfo{person}{Ning Shang},
  \bibinfo{person}{Yilei Yang}, \bibinfo{person}{Weijiang Xu},
  \bibinfo{person}{Mao Yang}, \bibinfo{person}{Lintao Zhang}, {and}
  \bibinfo{person}{Lidong Zhou}.} \bibinfo{year}{2024}\natexlab{}.
\newblock
  \showarticletitle{\href{https://www.usenix.org/conference/osdi24/presentation/lin-zhiqi}{{nnScaler:
  Constraint-Guided Parallelization Plan Generation for Deep Learning
  Training}}}. In \bibinfo{booktitle}{\emph{OSDI}}.
\newblock


\bibitem[Liu et~al\mbox{.}(2024b)]%
        {liu2024neuronabox}
\bibfield{author}{\bibinfo{person}{Banruo Liu},
  \bibinfo{person}{Mubarak~Adetunji Ojewale}, \bibinfo{person}{Yuhan Ding},
  {and} \bibinfo{person}{Marco Canini}.} \bibinfo{year}{2024}\natexlab{b}.
\newblock
  \showarticletitle{\href{https://doi.org/10.1145/3678015.3680478}{{Towards a
  Flexible and High-Fidelity Approach to Distributed DNN Training Emulation}}}.
  In \bibinfo{booktitle}{\emph{APSys}}.
\newblock


\bibitem[Liu et~al\mbox{.}(2024a)]%
        {liu2024teccl}
\bibfield{author}{\bibinfo{person}{Xuting Liu}, \bibinfo{person}{Behnaz
  Arzani}, \bibinfo{person}{Siva Kesava~Reddy Kakarla},
  \bibinfo{person}{Liangyu Zhao}, \bibinfo{person}{Vincent Liu},
  \bibinfo{person}{Miguel Castro}, \bibinfo{person}{Srikanth Kandula}, {and}
  \bibinfo{person}{Luke Marshall}.} \bibinfo{year}{2024}\natexlab{a}.
\newblock
  \showarticletitle{\href{https://doi.org/10.1145/3651890.3672249}{{Rethinking
  Machine Learning Collective Communication as a Multi-Commodity Flow
  Problem}}}. In \bibinfo{booktitle}{\emph{SIGCOMM}}.
\newblock


\bibitem[Mai et~al\mbox{.}(2020)]%
        {mai2020kungfu}
\bibfield{author}{\bibinfo{person}{Luo Mai}, \bibinfo{person}{Guo Li},
  \bibinfo{person}{Marcel Wagenl{\"a}nder}, \bibinfo{person}{Konstantinos
  Fertakis}, \bibinfo{person}{Andrei-Octavian Brabete}, {and}
  \bibinfo{person}{Peter Pietzuch}.} \bibinfo{year}{2020}\natexlab{}.
\newblock
  \showarticletitle{\href{https://www.usenix.org/conference/osdi20/presentation/mai}{{KungFu:
  Making Training in Distributed Machine Learning Adaptive}}}. In
  \bibinfo{booktitle}{\emph{OSDI}}.
\newblock


\bibitem[Markov et~al\mbox{.}(2024)]%
        {markov2024lgreco}
\bibfield{author}{\bibinfo{person}{Ilia Markov}, \bibinfo{person}{Kaveh Alim},
  \bibinfo{person}{Elias Frantar}, {and} \bibinfo{person}{Dan Alistarh}.}
  \bibinfo{year}{2024}\natexlab{}.
\newblock
  \showarticletitle{\href{https://proceedings.mlsys.org/paper_files/paper/2024/hash/9069a8976ff06f6443e7f4172990a580-Abstract-Conference.html}{{L-GreCo:
  Layerwise-adaptive Gradient Compression For Efficient Data-parallel Deep
  Learning}}}. In \bibinfo{booktitle}{\emph{MLSys}}.
\newblock


\bibitem[Maurya et~al\mbox{.}(2024)]%
        {maurya2024datastatesllm}
\bibfield{author}{\bibinfo{person}{Avinash Maurya}, \bibinfo{person}{Robert
  Underwood}, \bibinfo{person}{M.~Mustafa Rafique}, \bibinfo{person}{Franck
  Cappello}, {and} \bibinfo{person}{Bogdan Nicolae}.}
  \bibinfo{year}{2024}\natexlab{}.
\newblock
  \showarticletitle{\href{https://doi.org/10.1145/3625549.3658685}{{DataStates-LLM:
  Lazy Asynchronous Checkpointing for Large Language Models}}}. In
  \bibinfo{booktitle}{\emph{HPDC}}.
\newblock


\bibitem[{NVIDIA}(2024)]%
        {nccl}
\bibfield{author}{\bibinfo{person}{{NVIDIA}}.} \bibinfo{year}{2024}\natexlab{}.
\newblock \bibinfo{title}{NVIDIA Collective Communication Library (NCCL)}.
\newblock
\newblock
\urldef\tempurl%
\url{https://developer.nvidia.com/nccl}
\showURL{%
\tempurl}


\bibitem[Patke et~al\mbox{.}(2024)]%
        {patke2024qlm}
\bibfield{author}{\bibinfo{person}{Archit Patke}, \bibinfo{person}{Dhemath
  Reddy}, \bibinfo{person}{Saurabh Jha}, \bibinfo{person}{Haoran Qiu},
  \bibinfo{person}{Christian Pinto}, \bibinfo{person}{Chandra Narayanaswami},
  \bibinfo{person}{Zbigniew Kalbarczyk}, {and} \bibinfo{person}{Ravishankar
  Iyer}.} \bibinfo{year}{2024}\natexlab{}.
\newblock
  \showarticletitle{\href{https://doi.org/10.1145/3698038.3698523}{{Queue
  Management for SLO-Oriented Large Language Model Serving}}}. In
  \bibinfo{booktitle}{\emph{SoCC}}.
\newblock


\bibitem[Qian et~al\mbox{.}(2024)]%
        {qian2024alibabahpn}
\bibfield{author}{\bibinfo{person}{Kun Qian}, \bibinfo{person}{Yongqing Xi},
  \bibinfo{person}{Jiamin Cao}, \bibinfo{person}{Jiaqi Gao},
  \bibinfo{person}{Yichi Xu}, \bibinfo{person}{Yu Guan},
  \bibinfo{person}{Binzhang Fu}, \bibinfo{person}{Xuemei Shi},
  \bibinfo{person}{Fangbo Zhu}, \bibinfo{person}{Rui Miao},
  \bibinfo{person}{Chao Wang}, \bibinfo{person}{Peng Wang},
  \bibinfo{person}{Pengcheng Zhang}, \bibinfo{person}{Xianlong Zeng},
  \bibinfo{person}{Eddie Ruan}, \bibinfo{person}{Zhiping Yao},
  \bibinfo{person}{Ennan Zhai}, {and} \bibinfo{person}{Dennis Cai}.}
  \bibinfo{year}{2024}\natexlab{}.
\newblock
  \showarticletitle{\href{https://doi.org/10.1145/3651890.3672265}{{Alibaba
  HPN: A Data Center Network for Large Language Model Training}}}. In
  \bibinfo{booktitle}{\emph{SIGCOMM}}.
\newblock


\bibitem[Qiao et~al\mbox{.}(2021)]%
        {qiao2021pollux}
\bibfield{author}{\bibinfo{person}{Aurick Qiao}, \bibinfo{person}{Sang~Keun
  Choe}, \bibinfo{person}{Suhas~Jayaram Subramanya}, \bibinfo{person}{Willie
  Neiswanger}, \bibinfo{person}{Qirong Ho}, \bibinfo{person}{Hao Zhang},
  \bibinfo{person}{Gregory~R. Ganger}, {and} \bibinfo{person}{Eric~P. Xing}.}
  \bibinfo{year}{2021}\natexlab{}.
\newblock
  \showarticletitle{\href{https://www.usenix.org/conference/osdi21/presentation/qiao}{{Pollux:
  Co-adaptive Cluster Scheduling for Goodput-Optimized Deep Learning}}}. In
  \bibinfo{booktitle}{\emph{OSDI}}.
\newblock


\bibitem[Rajbhandari et~al\mbox{.}(2020)]%
        {rajbhandari2020zero}
\bibfield{author}{\bibinfo{person}{Samyam Rajbhandari}, \bibinfo{person}{Jeff
  Rasley}, \bibinfo{person}{Olatunji Ruwase}, {and} \bibinfo{person}{Yuxiong
  He}.} \bibinfo{year}{2020}\natexlab{}.
\newblock
  \showarticletitle{\href{https://doi.org/10.5555/3433701.3433727}{{ZeRO:
  Memory Optimizations Toward Training Trillion Parameter Models}}}. In
  \bibinfo{booktitle}{\emph{SC}}.
\newblock


\bibitem[Sculley et~al\mbox{.}(2015)]%
        {sculley2015technicaldebt}
\bibfield{author}{\bibinfo{person}{D. Sculley}, \bibinfo{person}{Gary Holt},
  \bibinfo{person}{Daniel Golovin}, \bibinfo{person}{Eugene Davydov},
  \bibinfo{person}{Todd Phillips}, \bibinfo{person}{Dietmar Ebner},
  \bibinfo{person}{Vinay Chaudhary}, \bibinfo{person}{Michael Young},
  \bibinfo{person}{Jean-Fran\c{c}ois Crespo}, {and} \bibinfo{person}{Dan
  Dennison}.} \bibinfo{year}{2015}\natexlab{}.
\newblock
  \showarticletitle{\href{https://proceedings.neurips.cc/paper_files/paper/2015/file/86df7dcfd896fcaf2674f757a2463eba-Paper.pdf}{{Hidden
  Technical Debt in Machine Learning Systems}}}. In
  \bibinfo{booktitle}{\emph{NeurIPS}}.
\newblock


\bibitem[Shoeybi et~al\mbox{.}(2020)]%
        {shoeybi2020megatronlm}
\bibfield{author}{\bibinfo{person}{Mohammad Shoeybi}, \bibinfo{person}{Mostofa
  Patwary}, \bibinfo{person}{Raul Puri}, \bibinfo{person}{Patrick LeGresley},
  \bibinfo{person}{Jared Casper}, {and} \bibinfo{person}{Bryan Catanzaro}.}
  \bibinfo{year}{2020}\natexlab{}.
\newblock \bibinfo{title}{\href{https://arxiv.org/abs/1909.08053}{{Megatron-LM:
  Training Multi-Billion Parameter Language Models Using Model Parallelism}}}.
\newblock
\newblock
\showeprint[arxiv]{1909.08053}~[cs.CL]


\bibitem[Stojkovic et~al\mbox{.}(2025)]%
        {stojkovic2024dynamollm}
\bibfield{author}{\bibinfo{person}{Jovan Stojkovic}, \bibinfo{person}{Chaojie
  Zhang}, \bibinfo{person}{Íñigo Goiri}, \bibinfo{person}{Josep Torrellas},
  {and} \bibinfo{person}{Esha Choukse}.} \bibinfo{year}{2025}\natexlab{}.
\newblock
  \showarticletitle{\href{https://doi.org/10.1109/HPCA61900.2025.00102}{{DynamoLLM:
  Designing LLM Inference Clusters for Performance and Energy Efficiency}}}. In
  \bibinfo{booktitle}{\emph{HPCA}}.
\newblock


\bibitem[Tarnawski et~al\mbox{.}(2021)]%
        {tarnawski2021piper}
\bibfield{author}{\bibinfo{person}{Jakub~M. Tarnawski}, \bibinfo{person}{Deepak
  Narayanan}, {and} \bibinfo{person}{Amar Phanishayee}.}
  \bibinfo{year}{2021}\natexlab{}.
\newblock
  \showarticletitle{\href{https://proceedings.neurips.cc/paper_files/paper/2021/file/d01eeca8b24321cd2fe89dd85b9beb51-Paper.pdf}{{Piper:
  Multidimensional Planner for DNN Parallelization}}}. In
  \bibinfo{booktitle}{\emph{NeurIPS}}.
\newblock


\bibitem[Tay(2024)]%
        {reka-llms-wilderness}
\bibfield{author}{\bibinfo{person}{Yi Tay}.} \bibinfo{year}{2024}\natexlab{}.
\newblock \bibinfo{title}{Training Great LLMs Entirely from Ground Zero in the
  Wilderness}.
\newblock
\newblock
\urldef\tempurl%
\url{https://www.yitay.net/blog/training-great-llms-entirely-from-ground-zero-in-the-wilderness}
\showURL{%
\tempurl}


\bibitem[Wang et~al\mbox{.}(2024)]%
        {wang2024fastpersist}
\bibfield{author}{\bibinfo{person}{Guanhua Wang}, \bibinfo{person}{Olatunji
  Ruwase}, \bibinfo{person}{Bing Xie}, {and} \bibinfo{person}{Yuxiong He}.}
  \bibinfo{year}{2024}\natexlab{}.
\newblock \bibinfo{title}{\href{https://arxiv.org/abs/2406.13768}{{FastPersist:
  Accelerating Model Checkpointing in Deep Learning}}}.
\newblock
\newblock
\showeprint[arxiv]{2406.13768}~[cs.DC]


\bibitem[Wang et~al\mbox{.}(2019)]%
        {wang2019tofu}
\bibfield{author}{\bibinfo{person}{Minjie Wang}, \bibinfo{person}{Chien-chin
  Huang}, {and} \bibinfo{person}{Jinyang Li}.} \bibinfo{year}{2019}\natexlab{}.
\newblock
  \showarticletitle{\href{https://doi.org/10.1145/3302424.3303953}{{Supporting
  Very Large Models using Automatic Dataflow Graph Partitioning}}}. In
  \bibinfo{booktitle}{\emph{EuroSys}}.
\newblock


\bibitem[Wang et~al\mbox{.}(2023)]%
        {wang2023gemini}
\bibfield{author}{\bibinfo{person}{Zhuang Wang}, \bibinfo{person}{Zhen Jia},
  \bibinfo{person}{Shuai Zheng}, \bibinfo{person}{Zhen Zhang},
  \bibinfo{person}{Xinwei Fu}, \bibinfo{person}{T.~S.~Eugene Ng}, {and}
  \bibinfo{person}{Yida Wang}.} \bibinfo{year}{2023}\natexlab{}.
\newblock
  \showarticletitle{\href{https://doi.org/10.1145/3600006.3613145}{{GEMINI:
  Fast Failure Recovery in Distributed Training with In-Memory Checkpoints}}}.
  In \bibinfo{booktitle}{\emph{SOSP}}.
\newblock


\bibitem[Wolf et~al\mbox{.}(2020)]%
        {wolf2020transformers}
\bibfield{author}{\bibinfo{person}{Thomas Wolf}, \bibinfo{person}{Lysandre
  Debut}, \bibinfo{person}{Victor Sanh}, \bibinfo{person}{Julien Chaumond},
  \bibinfo{person}{Clement Delangue}, \bibinfo{person}{Anthony Moi},
  \bibinfo{person}{Pierric Cistac}, \bibinfo{person}{Tim Rault},
  \bibinfo{person}{Rémi Louf}, \bibinfo{person}{Morgan Funtowicz},
  \bibinfo{person}{Joe Davison}, \bibinfo{person}{Sam Shleifer},
  \bibinfo{person}{Patrick von Platen}, \bibinfo{person}{Clara Ma},
  \bibinfo{person}{Yacine Jernite}, \bibinfo{person}{Julien Plu},
  \bibinfo{person}{Canwen Xu}, \bibinfo{person}{Teven~Le Scao},
  \bibinfo{person}{Sylvain Gugger}, \bibinfo{person}{Mariama Drame},
  \bibinfo{person}{Quentin Lhoest}, {and} \bibinfo{person}{Alexander~M. Rush}.}
  \bibinfo{year}{2020}\natexlab{}.
\newblock
  \showarticletitle{\href{https://www.aclweb.org/anthology/2020.emnlp-demos.6}{{Transformers:
  State-of-the-Art Natural Language Processing}}}. In
  \bibinfo{booktitle}{\emph{EMNLP: System Demonstrations}}.
\newblock


\bibitem[Wu et~al\mbox{.}(2024a)]%
        {wu2024sustainableai}
\bibfield{author}{\bibinfo{person}{Carole-Jean Wu}, \bibinfo{person}{Bilge
  Acun}, \bibinfo{person}{Ramya Raghavendra}, {and} \bibinfo{person}{Kim
  Hazelwood}.} \bibinfo{year}{2024}\natexlab{a}.
\newblock
  \showarticletitle{\href{https://doi.org/10.1109/MM.2024.3409275}{{Beyond
  Efficiency: Scaling AI Sustainably}}}.
\newblock \bibinfo{journal}{\emph{IEEE Micro}} \bibinfo{volume}{44},
  \bibinfo{number}{5} (\bibinfo{year}{2024}).
\newblock


\bibitem[Wu et~al\mbox{.}(2022)]%
        {wu2022sustainableai}
\bibfield{author}{\bibinfo{person}{Carole-Jean Wu}, \bibinfo{person}{Ramya
  Raghavendra}, \bibinfo{person}{Udit Gupta}, \bibinfo{person}{Bilge Acun},
  \bibinfo{person}{Newsha Ardalani}, \bibinfo{person}{Kiwan Maeng},
  \bibinfo{person}{Gloria Chang}, \bibinfo{person}{Fiona Aga},
  \bibinfo{person}{Jinshi Huang}, \bibinfo{person}{Charles Bai},
  \bibinfo{person}{Michael Gschwind}, \bibinfo{person}{Anurag Gupta},
  \bibinfo{person}{Myle Ott}, \bibinfo{person}{Anastasia Melnikov},
  \bibinfo{person}{Salvatore Candido}, \bibinfo{person}{David Brooks},
  \bibinfo{person}{Geeta Chauhan}, \bibinfo{person}{Benjamin Lee},
  \bibinfo{person}{Hsien-Hsin Lee}, \bibinfo{person}{Bugra Akyildiz},
  \bibinfo{person}{Maximilian Balandat}, \bibinfo{person}{Joe Spisak},
  \bibinfo{person}{Ravi Jain}, \bibinfo{person}{Mike Rabbat}, {and}
  \bibinfo{person}{Kim Hazelwood}.} \bibinfo{year}{2022}\natexlab{}.
\newblock
  \showarticletitle{\href{https://proceedings.mlsys.org/paper_files/paper/2022/file/462211f67c7d858f663355eff93b745e-Paper.pdf}{{Sustainable
  AI: Environmental Implications, Challenges and Opportunities}}}. In
  \bibinfo{booktitle}{\emph{MLSys}}.
\newblock


\bibitem[Wu et~al\mbox{.}(2024b)]%
        {wu2024mccs}
\bibfield{author}{\bibinfo{person}{Yongji Wu}, \bibinfo{person}{Yechen Xu},
  \bibinfo{person}{Jingrong Chen}, \bibinfo{person}{Zhaodong Wang},
  \bibinfo{person}{Ying Zhang}, \bibinfo{person}{Matthew Lentz}, {and}
  \bibinfo{person}{Danyang Zhuo}.} \bibinfo{year}{2024}\natexlab{b}.
\newblock
  \showarticletitle{\href{https://doi.org/10.1145/3651890.3672252}{{MCCS: A
  Service-based Approach to Collective Communication for Multi-Tenant Cloud}}}.
  In \bibinfo{booktitle}{\emph{SIGCOMM}}.
\newblock


\bibitem[Xie et~al\mbox{.}(2008)]%
        {xie2008p4p}
\bibfield{author}{\bibinfo{person}{Haiyong Xie}, \bibinfo{person}{Y.~Richard
  Yang}, \bibinfo{person}{Arvind Krishnamurthy}, \bibinfo{person}{Yanbin~Grace
  Liu}, {and} \bibinfo{person}{Abraham Silberschatz}.}
  \bibinfo{year}{2008}\natexlab{}.
\newblock
  \showarticletitle{\href{https://doi.org/10.1145/1402958.1402999}{{P4P:
  Provider Portal for Applications}}}. In \bibinfo{booktitle}{\emph{SIGCOMM}}.
\newblock


\bibitem[Xin et~al\mbox{.}(2023)]%
        {xin2023kimad}
\bibfield{author}{\bibinfo{person}{Jihao Xin}, \bibinfo{person}{Ivan Ilin},
  \bibinfo{person}{Shunkang Zhang}, \bibinfo{person}{Marco Canini}, {and}
  \bibinfo{person}{Peter Richt\'{a}rik}.} \bibinfo{year}{2023}\natexlab{}.
\newblock
  \showarticletitle{\href{https://doi.org/10.1145/3630048.3630184}{{Kimad:
  Adaptive Gradient Compression with Bandwidth Awareness}}}. In
  \bibinfo{booktitle}{\emph{DistributedML}}.
\newblock


\bibitem[Xiong et~al\mbox{.}(2024)]%
        {xiong2024superbench}
\bibfield{author}{\bibinfo{person}{Yifan Xiong}, \bibinfo{person}{Yuting
  Jiang}, \bibinfo{person}{Ziyue Yang}, \bibinfo{person}{Lei Qu},
  \bibinfo{person}{Guoshuai Zhao}, \bibinfo{person}{Shuguang Liu},
  \bibinfo{person}{Dong Zhong}, \bibinfo{person}{Boris Pinzur},
  \bibinfo{person}{Jie Zhang}, \bibinfo{person}{Yang Wang},
  \bibinfo{person}{Jithin Jose}, \bibinfo{person}{Hossein Pourreza},
  \bibinfo{person}{Jeff Baxter}, \bibinfo{person}{Kushal Datta},
  \bibinfo{person}{Prabhat Ram}, \bibinfo{person}{Luke Melton},
  \bibinfo{person}{Joe Chau}, \bibinfo{person}{Peng Cheng},
  \bibinfo{person}{Yongqiang Xiong}, {and} \bibinfo{person}{Lidong Zhou}.}
  \bibinfo{year}{2024}\natexlab{}.
\newblock
  \showarticletitle{\href{https://www.usenix.org/conference/atc24/presentation/xiong}
  {{SuperBench: Improving Cloud AI Infrastructure Reliability with Proactive
  Validation}}}. In \bibinfo{booktitle}{\emph{USENIX ATC}}.
\newblock


\bibitem[You et~al\mbox{.}(2023)]%
        {you2023zeus}
\bibfield{author}{\bibinfo{person}{Jie You}, \bibinfo{person}{Jae-Won Chung},
  {and} \bibinfo{person}{Mosharaf Chowdhury}.} \bibinfo{year}{2023}\natexlab{}.
\newblock
  \showarticletitle{\href{https://www.usenix.org/conference/nsdi23/presentation/you}{{Zeus:
  Understanding and Optimizing GPU Energy Consumption of DNN Training}}}. In
  \bibinfo{booktitle}{\emph{NSDI}}.
\newblock


\bibitem[Zhao et~al\mbox{.}(2024)]%
        {zhao2024serverless}
\bibfield{author}{\bibinfo{person}{Yuxuan Zhao}, \bibinfo{person}{Weikang
  Weng}, \bibinfo{person}{Rob van Nieuwpoort}, {and} \bibinfo{person}{Alexandru
  Uta}.} \bibinfo{year}{2024}\natexlab{}.
\newblock \showarticletitle{\href{https://doi.org/10.1145/3652892.3700757}{{In
  Serverless, OS Scheduler Choice Costs Money: A Hybrid Scheduling Approach for
  Cheaper FaaS}}}. In \bibinfo{booktitle}{\emph{Middleware}}.
\newblock


\bibitem[Zheng et~al\mbox{.}(2022)]%
        {zheng2022alpa}
\bibfield{author}{\bibinfo{person}{Lianmin Zheng}, \bibinfo{person}{Zhuohan
  Li}, \bibinfo{person}{Hao Zhang}, \bibinfo{person}{Yonghao Zhuang},
  \bibinfo{person}{Zhifeng Chen}, \bibinfo{person}{Yanping Huang},
  \bibinfo{person}{Yida Wang}, \bibinfo{person}{Yuanzhong Xu},
  \bibinfo{person}{Danyang Zhuo}, \bibinfo{person}{Eric~P. Xing},
  \bibinfo{person}{Joseph~E. Gonzalez}, {and} \bibinfo{person}{Ion Stoica}.}
  \bibinfo{year}{2022}\natexlab{}.
\newblock
  \showarticletitle{\href{https://www.usenix.org/conference/osdi22/presentation/zheng-lianmin}{{Alpa:
  Automating Inter- and Intra-Operator Parallelism for Distributed Deep
  Learning}}}. In \bibinfo{booktitle}{\emph{OSDI}}.
\newblock


\end{thebibliography}

\end{document}